\documentclass[11pt]{article}

\usepackage{graphicx}

\def\lromn#1{\uppercase\expandafter{\romannumeral#1}}

\begin{document}

\begin{flushright}
\today \\
\end{flushright}

\begin{center}
\begin{Large}

{\bf Basic oscillation measurables  in 
the neutrino  pair beam 
}
\end{Large}

\vspace{2cm}
T.~Asaka,  M. Tanaka$^{\dagger}$ and
M.~Yoshimura$^{\ddagger}$

\vspace{0.5cm}

Department of Physics, Niigata University, 950-2181 Niigata, Japan
\\[2ex]
$^{\dagger}$
Department of Physics, Graduate School of Science, \\
             Osaka University, Toyonaka, Osaka 560-0043, Japan
 \\[2ex]

$^{\ddagger}$
Center of Quantum Universe, Faculty of
Science, Okayama University \\
Tsushima-naka 3-1-1 Kita-ku Okayama
700-8530 Japan
\end{center}

\vspace{3cm}

\begin{center}
\begin{Large}
{\bf ABSTRACT}
\end{Large}
\end{center}

It is shown that the vector current contribution
of neutrino interaction with electrons in ion
gives rise to oscillating component,
which is absent for the axial-vector contribution, when
a single neutrino is detected in the recently
proposed neutrino pair beam.
CP violation measurements are thus
possible with high precision along with
determination of mass hierarchical patterns.

\vspace{4cm}
PACS numbers
\hspace{0.5cm} 
13.15.+g, 
14.60.Pq, 

Keywords
\hspace{0.5cm} 
CP violation,
CP-even neutrino pair beam, 
heavy ion synchrotron
\newpage

{\bf Introduction} \hspace{0.3cm} A new strong source of neutrinos
consisting of all flavor pairs of $\nu_a$ and $\bar{\nu}_a$ ($a=
e,\mu, \tau$) was recently proposed to further accelerate neutrino
physics experiments~\cite{pair beam}.  It was however pointed out
in~\cite{n-oscillation in pair beam} that a single neutrino detection
in the neutrino pair beam does not exhibit oscillation pattern, making
detection of oscillation harder due to large backgrounds in double
event detection.  This disappearance of oscillation is based on (i)
the unitarity of the $3\times 3$ neutrino mixing matrix and (ii) the
equality of pair emission amplitude squared that holds for the dominant axial
vector contribution of light ions.

In the present work we show that the second condition, the equality of
pair emission amplitude squared, does not hold in the vector current
contribution (sub-dominant, when ionic electrons move with
non-relativistic velocities, but may be comparable to the axial vector
contribution in heavy ions) of pair emission amplitude, hence the
emergence of oscillation patterns occurs from the vector contribution.
When neutrino oscillation is made possible this way, the CP violating
(CPV) parameter determination (the CPV phase $\delta$ common to both
Dirac and Majorana cases) becomes possible.

We derive basic formulas for the three-flavor neutrino scheme
including the earth matter effect 
and present numerical outputs of quantities for new
experiments using the neutrino pair beam.
It is found that oscillation patterns appear in all $\nu_a, \bar{\nu}_a\,, a= e, \mu, \tau$,
but determination of CPV parameter is possible only by detection of 
$\nu_{\mu}, \bar{\nu}_{\mu}$ and tau-neutrinos.
Electron neutrinos do not allow  CPV determination.
If the accelerator ring is placed in the underground of depth $d$,
the neutrino pair beam appear on earth at a distance $\sim \sqrt{2 d R}$
with $R$ the radius of the earth.
This distance is $\sim 35$ km for $d=100$ m.
It is found that one can do sensitive CPV measurements at
this distance.

Throughout this work we use the natural unit of $\hbar = c = 1$.

{\bf How oscillation pattern appears in single neutrino detection}
\hspace{0.3cm} We shall follow notations of \cite{n-oscillation in
  pair beam}.  The probability amplitude of the entire process
consists of three parts: the production, the propagation, and the
detection due to charged current interaction (neutral current
interaction is much smaller, hence not considered here), each to be
multiplied at the amplitude level.  Thus, one may write the
probability for the $\nu_a$ neutrino quasi-elastic scattering (with
$J^\alpha$ the nucleon weak current) as
\begin{eqnarray}
  &&
  \sum_{c}
  \left(\frac{G_F}{\sqrt{2}}\right)^2 
  \bar{\nu}_a \gamma_{\alpha} ( 1- \gamma_5) l_a J^{\alpha} \,
  \bar{l}_a \gamma_{\beta} ( 1- \gamma_5) \nu_a  (J^{\beta})^{\dagger} \,
  \left|
    \sum_b
    \langle \bar{c} | e^{-i \bar H \bar L} | \bar{b} \rangle
    \langle a | e^{-i H L} | b \rangle  
    {\cal M}_{\bar{b}b}(1,2)
  \right|^2
  \,,
\end{eqnarray}
where  $H$ ($\bar H$) is the hamiltonian for propagation of neutrino (antineutrino) including
earth-induced matter effect \cite{wolfenstein}, \cite{barger etc}, \cite{xing}, which is in the flavor basis
\begin{eqnarray}
&&
H = U^* \left(
\begin{array}{ccc}
\displaystyle
\frac{m_1^2} {2E} & 0  & 0\\
0 & \displaystyle \frac{m_2^2} {2E}  & 0\\
0 & 0  & \displaystyle \frac{m_3^2} {2E}
\end{array}
\right) U^T
+ \sqrt{2} G_F n_e \left(
\begin{array}{ccc}
1 & 0  & 0\\
0 &  0 & 0\\
0 & 0  &0
\end{array}
\right)
\,,
\label {hamiltonian with earth matter}
\end{eqnarray}
where $U_{a i}$ is the neutrino mixing matrix with $|a \rangle =
\sum_i U_{ai}^* |i \rangle, a = e, \mu, \tau, i = 1,2,3$, and $n_e$ is
the number density of electrons in the earth.  $\bar H$ can be obtained by
replacing $U \to U^\ast$ and changing the sign in the second term
$\propto G_F$.

We shall denote three eigenvalues by $\lambda_i$ for neutrinos, and
$\bar{\lambda}_i$ for anti-neutrinos.  Let $V (\sim U)$ and $\bar{V}$
are unitary $3\times 3$ matrices that diagonalize the hamiltonian $H$
for neutrino and $\bar{H}$ for anti-neutrino, including the earth
matter effect.  The propagation amplitude is then
\begin{eqnarray}
&&
\langle a | e^{-iH L} | b \rangle = 
\sum_i V_{ai} V_{bi}^* e^{-i\lambda_i L}
\,, \hspace{0.5cm}
\langle \bar{c} | e^{-i \bar H \bar L} | \bar{b} \rangle = \sum_i \bar{V}_{ci}^* \bar{V}_{bi} e^{-i \bar{\lambda}_i \bar L}
\,,
\\ &&
\sum_b
\langle \bar{c} | e^{-i \bar H L} | \bar{b} \rangle  \langle a | e^{-iH L} | b \rangle c_b 
=
\sum_{ij} V_{ai} \bar{V}_{c j}^\ast \xi_{ij} 
\exp[-i ( \lambda_i L + \bar{\lambda}_j \bar L)]
\,,
\hspace{0.5cm}
\xi_{ij} =  
\sum_b c_b \, V_{bi}^\ast \bar{V}_{b j}
\,, 
\,.
\end{eqnarray}
The factor $c_b$ arises from the production amplitude ${\cal
  M}_{\bar{b}b}(1,2)$ and it is $(c_b^A )= \frac{1}{2} (1,-1,-1)$ for
the axial vector contribution and for the vector contribution,
\begin{eqnarray}
(c_b^V) = 
\left( \frac{1}{2}(1+ 4 \sin^2 \theta_w)\,,
  - \frac{1}{2}(1- 4 \sin^2 \theta_w)\,,
- \frac{1}{2}(1- 4 \sin^2 \theta_w) \right)
\,,
\end{eqnarray}
with the weak mixing angle $\theta_w$.
The precise relation between neutrino and anti-neutrino eigenvalue problem is given by
\begin{eqnarray}
&&
\bar{\lambda}(G_F) = \lambda(-G_F)
\,, \hspace{0.5cm}
\bar{V}_{ai} (G_F) = V_{ai} (-G_F)
\,.
\end{eqnarray}
The rate of neutrino $\nu_a$ detected and $\bar{\nu}_{c}$ undetected
contains the squared propagation factor,
\begin{eqnarray}
  &&
  \sum_c 
  \Bigl| 
    \sum_{ij} V_{ai} \bar{V}_{c j}^\ast \xi_{ij} 
    \exp[-i ( \lambda_i L + \bar{\lambda}_j \bar L)] 
  \Bigr|^2 
  =
  \sum_{ijkl} 
  \sum_c V_{ai}  V_{ak}^* \bar{V}_{c j}^\ast  \bar{V}_{c l}
  \xi_{ij} \xi_{kl}^* 
  \exp [ - i ( \lambda_i - \lambda_k ) L]
  \exp [ - i ( \bar{\lambda}_j - \bar{\lambda}_l ) \bar L]
\nonumber \\ &&
 = 
 \sum_{ik} V_{a i}  V_{a k}^*   p_{ik}
  \exp [ - i ( \lambda_i - \lambda_k ) L]
  \,, \hspace{0.5cm}
p_{ik} =
\sum_j \xi_{ij} \xi_{kj}^*
\,.
\end{eqnarray}
When $(|c_b^A|^2)=  (1,1,1)/4 \propto 1 $  for the axial vector contribution,
$p_{jl} = \delta_{jl}/4$ and 
the detection probability becomes 1/4, hence
no oscillation pattern exits.

The relevant weak amplitude for the vector part
gives oscillating components.
Candidate ions for circulation that contribute to
the vector current interaction are
Be-like heavy ions of  $2p 2s ^3P_1^-$
and Ne-like heavy ions of $2p^+ 3s ^3P_1^-$ (electron-hole system).

 {\bf Basic measurable quantities in neutrino pair beam}
\hspace{0.3cm}
We first note
\begin{eqnarray}
  &&
  p_{ik} 
  =
  \sum_j \xi_{ij} \xi_{kj}^*
  =
  \sum_b |c_b^V|^2 V_{b i}^\ast V_{b k}
  =
  \frac{1}{4}(1+ 4 \sin^2 \theta_w)^2 \,
  V_{e i}^*  V_{e k}
  + \frac{1}{4}(1- 4 \sin^2 \theta_w)^2 \,
  ( V_{\mu i}^*  V_{\mu k} + V_{\tau i}^*  V_{\tau k})
  \nonumber \\
  &&
  =
  \frac{1}{4}(1- 4 \sin^2 \theta_w)^2  \delta_{ik}
  + 4 \sin^2 \theta_W V_{ei}^\ast V_{ek}
  \,.
\end{eqnarray}
The detection probability of $\nu_a$ (when the other neutrino of the
pair is undetected) is given by the oscillation formula based on the
vector part of weak current,
\begin{eqnarray}
&&
P_a(E, L; m_i, \delta) \equiv 
\frac{1 }{ 3(1- 4 \sin^2 \theta_w)^2/4 + 4 \sin^2 \theta_w}
\nonumber \\ && \times
\left(
\frac{1}{4}(1- 4 \sin^2 \theta_w)^2
+ 4 \sin^2 \theta_w  | \sum_j U_{e j}^*U_{a j} \exp[-i   \frac{m^2_{j} L}{2E}]|^2
\right)
\,,
\label {normalized probability approximate}
\end{eqnarray}
with $\sin^2 \theta_w \sim 0.231$.
The formula (\ref{normalized probability approximate}) is valid when
the earth matter effect is neglected. 
When the earth matter effect is included,
one replaces $U \rightarrow V\,, m_j^2/2E \rightarrow \lambda_j$.
The quantity
$P_a(E, L; m_i, \delta)$ is the normalized probability: $\sum_a P_a(E, L; m_i, \delta) =1$.
The oscillating component in eq.~(\ref{normalized probability approximate}) 
is equivalent to the $\nu_e \rightarrow \nu_{a}\,, a= \mu, \tau$ appearance probability
multiplied by
\begin{eqnarray}
&&
\frac{4 \sin^2 \theta_w }{ 3(1- 4 \sin^2 \theta_w)^2/4 + 4 \sin^2 \theta_w} \sim 0.995
\,.
\end{eqnarray}
Thus, the constant off-set term $\propto (1- 4 \sin^2 \theta_w)^2$
in eq.~(\ref{normalized probability approximate}) is very small.
In the limit of $\sin^2 \theta_w = 1/4$ there is no contribution
to the vector part from Z-boson exchange.
Due to the dominance of $\nu_e \rightarrow \nu_a, a=\mu, \tau$
in the oscillating term, oscillation patterns in the pair beam
have similarities to the $\beta$ \cite{beta beam}
and $\beta^{\pm}$ beam \cite{beta pm beam}.

The most striking feature of the neutrino pair beam
is that circulating quantum ions produce coherent pairs
of all flavors, $\nu_a \bar{\nu}_a\,, a= e,\nu, \tau$.
When these pairs propagate, all mass eigen-states
get involved, and relevant oscillation
extrema at $L/E = 2\pi/\delta m^2_{ij}\,, (ij) = (12), (23), (13)$
may become relevant, making short baseline experiments
a feasible approach.

\begin{figure*}[htbp]
  \begin{center}
    \centerline{
      \includegraphics[width=9cm]{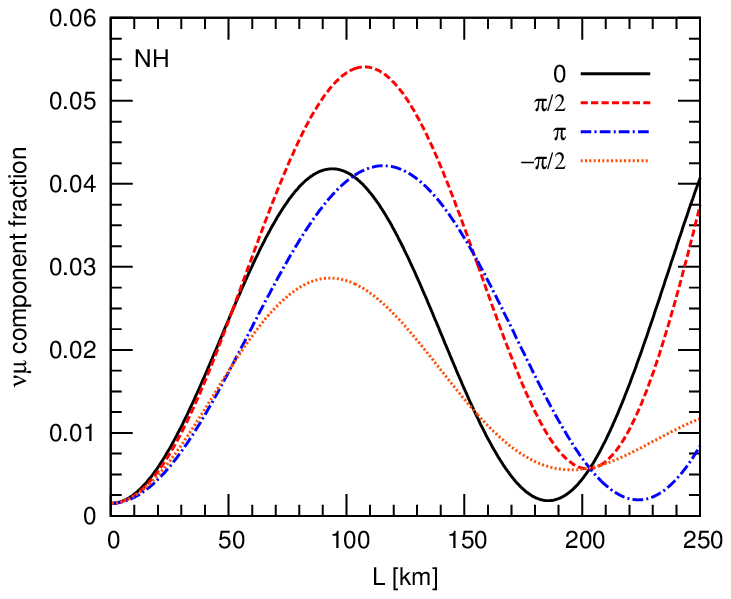}%
      \includegraphics[width=9cm]{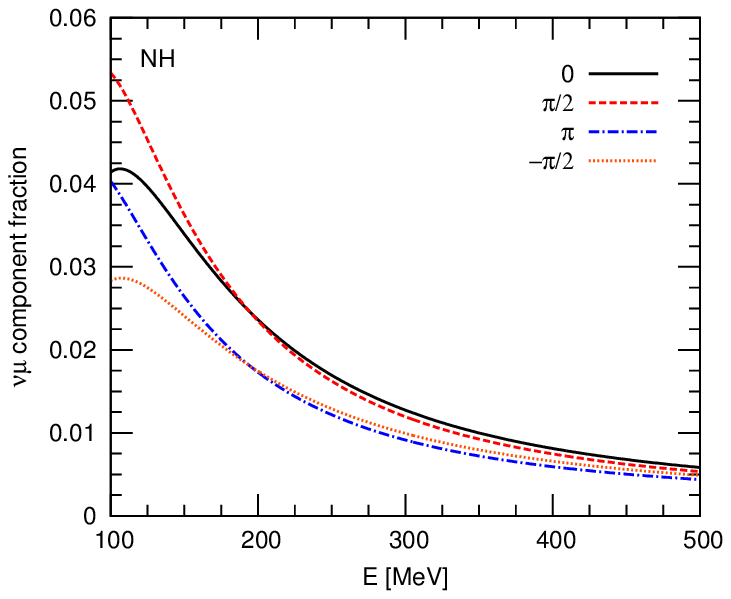}%
    }%
    \caption{$\nu_{\mu}$ component fraction of the pair beam, calculated by using
eq.~(\ref {normalized probability approximate}).
A few choices of CPV parameter $\delta$ are taken:
$0$ in solid black, $\pi/2$ in dashed red, $\pi$ in dash-dotted blue,
and $-\pi/2$ in dotted orange.
In the left panel the neutrino energy is fixed at 200 MeV.
In the right panel the distance is 50 km away from the ring
and the lowest $\nu_{\mu}$ energy should be set at $\sim$ 200 MeV to avoid
$\nu_{\mu} \rightarrow \mu$ threshold effect.
}
   \label{mu survival vs distance}
 \end{center} 
\end{figure*}

\begin{figure*}[htbp]
 \begin{center}
  \includegraphics[width=10cm]{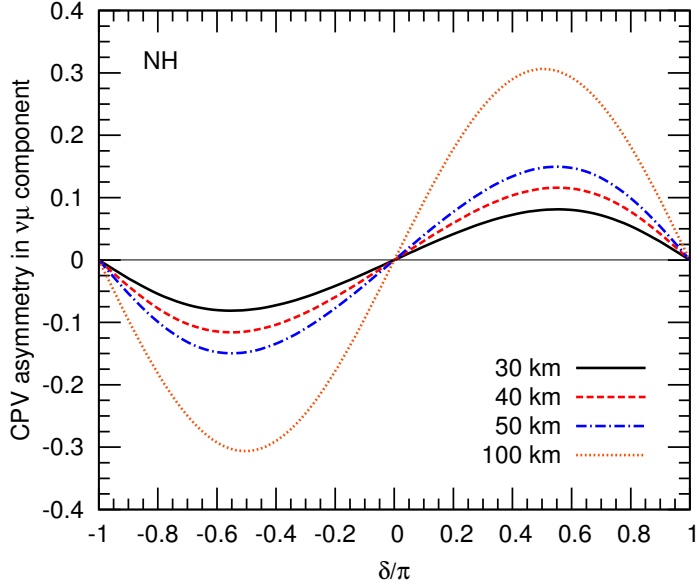}%
   \caption{$\nu_{\mu}$ CPV asymmetry plotted
against CPV parameter $\delta$ given by eq.~(\ref{cpv asymmetry}).
Assumed parameters are 
the neutrino energy 200 MeV,
the distance = 30 km in solid black, 40 km in dashed red,
50 km in dash-dotted blue, and 100 km in dotted orange.
The earth matter effect is negligible in these distances
as shown in Fig.~\ref{earth effect 2}.
}
   \label {mu asymmetry}
 \end{center} 
\end{figure*}

We illustrate numerical results of oscillation patterns and
CPV asymmetry in Figs.~\ref {mu survival vs distance}
and~\ref {mu asymmetry}, respectively.
We used neutrino data as determined
from neutrino oscillation experiments
\cite{mixing parameters}.
CPV asymmetry here is defined by
\begin{eqnarray}
&&
A_a(E, L; m_i, \delta) = \frac{P_a(E, L; m_i, \delta) - P_a(E, L; m_i, -\delta) }
{ P_a(E, L; m_i, \delta) + P_a(E, L; m_i, -\delta)}
\,.
\label {cpv asymmetry}
\end{eqnarray}
Experimentally, this quantity may be derived from measurements of
both $\nu_a$ and $\bar{\nu}_a$ events.

\begin{figure*}[htbp]
 \begin{center}
   \centerline{
  \includegraphics[width=9cm]{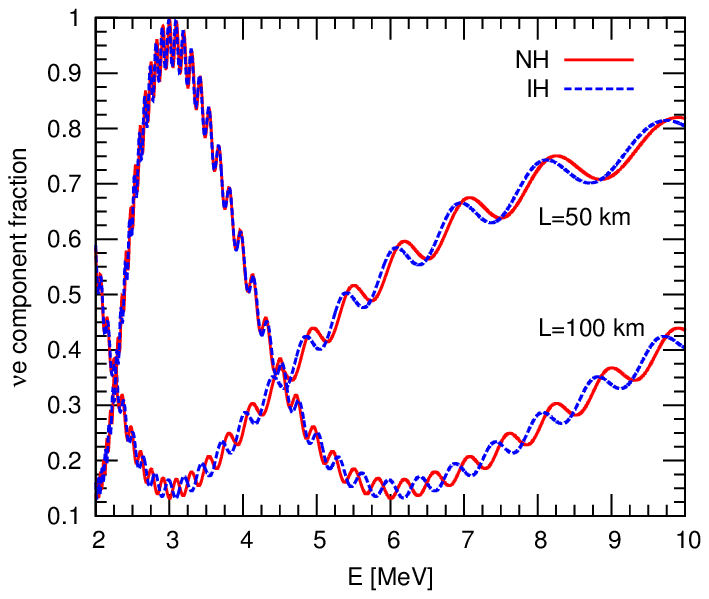}%
  \includegraphics[width=9cm]{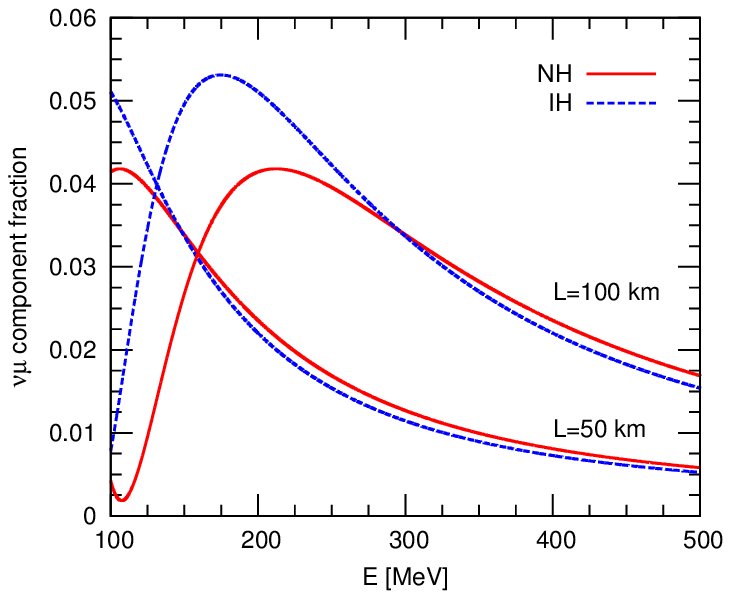}%
  }
  \caption{$\nu_{e}$ (left) and $\nu_\mu$ (right) component fractions of the pair beam, calculated by using
eq.~(\ref{normalized probability approximate})
for NH (solid red) and IH (dash-dotted blue).
The distances are 50 km and 100 km away from the ring.
}
   \label{e survival vs energy}
 \end{center} 
\end{figure*}

We note a few important results:
(i) CPV $\delta$ measurement is impossible for
$\nu_e, \bar{\nu}_e$ events, because the oscillation probability
appearing in eq.~(\ref{normalized probability approximate})
depends on the quantity $|U_{ej}|^2$, hence is insensitive to
$\delta$.
(ii) CPV asymmetry measured by detection of
$\nu_{\mu}, \bar{\nu}_{\mu}$ is small at the ion ring site due to
the unitarity relation $\sum_j U_{e j}^* U_{\mu j} = 0$ valid at small $L$. 
(iii)
Interestingly, as shown in Fig.~\ref{mu asymmetry},
CPV asymmetry of ${\cal O}(0.1)$ can be obtained even if the distance is
$L \sim 40$~km for $E = 200$ MeV.
(iv) The determination of the mass hierarchical pattern, normal or inverted,
namely NH/IH distinction is possible in the $\nu_e$ and $\nu_\mu$ components
as seen in Fig.~\ref{e survival vs energy}.
The advantage of $\nu_e$ component
was also pointed out in the reactor neutrino
experiment~\cite{petkov-piai}.

Comparison with Z-decay pair may be of interest.
In this case $(c_b^V) = - \frac{1}{2}(1- 4 \sin^2 \theta_w)(1,1,1)$.
Hence, the neutrino-pair beam from Z-boson decay
does not show the oscillation pattern when only
one neutrino is detected
\cite{smirnov-zatsepin}.

\begin{figure*}[htbp]
 \begin{center}
  \includegraphics[width=10cm]{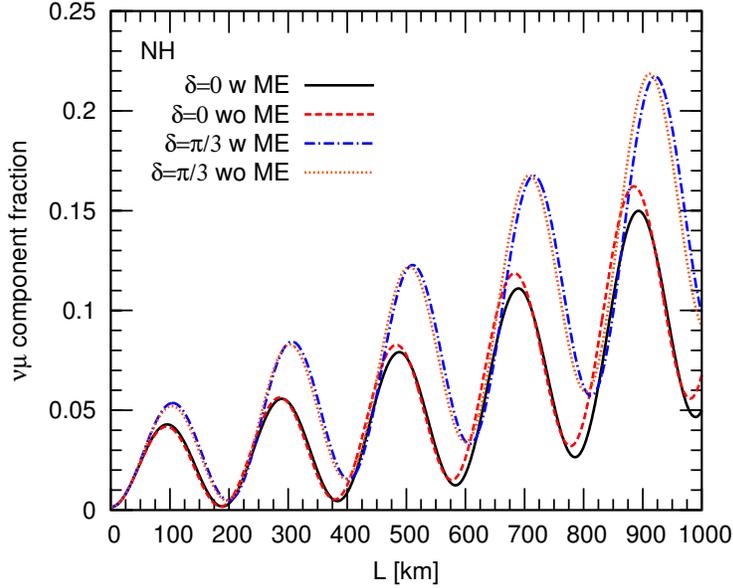}%
  \caption{$\nu_{\mu}$ oscillation pattern with and without the earth
    matter effect (ME).  The neutrino energy is fixed at 200 MeV.
    $\delta =0$ with ME in solid black, without ME
    in dashed red, $\delta = \pi/3$ with ME
    in dash-dotted blue, and without ME in dotted orange.  }
   \label {earth effect 1}
 \end{center} 
\end{figure*}

\begin{figure*}[htbp]
 \begin{center}
  \includegraphics[width=10cm]{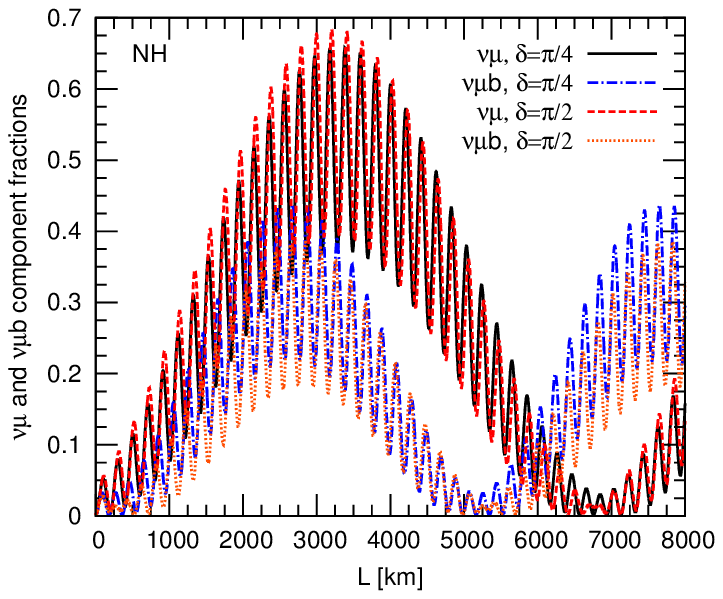}%
  \caption{$\nu_{\mu}$ oscillation pattern with the earth matter
    effect. The neutrino energy is fixed at 200 MeV.  $\delta =\pi/4$
    neutrino ($\nu_\mu$) in solid black, anti-neutrino ($\bar \nu_\mu=\nu_\mu$b) 
    in dash-dotted blue,
    $\delta = \pi/2$ neutrino in dashed red, and anti-neutrino in
    dotted orange.  }
   \label {earth effect 2}
 \end{center} 
\end{figure*}

To incorporate the earth matter effect,
we numerically diagonalize the effective hamiltonian (\ref{hamiltonian with earth matter})
\cite{sign of earth matter effect}.
The oscillation patterns including the earth matter effect are illustrated in
Figs.~\ref {earth effect 1} and \ref {earth effect 2}.
We took a pure  SiO$_2$ model with density $2.8$ g/cm$^3$ for the earth matter.

{\bf Rates of neutrino-pair production from quantum ion beam}
\hspace{0.3cm}
Calculations in \cite{n-oscillation in pair beam} 
are for the axial-vector contribution.
We now repeat calculations based on
the vector contribution, following \cite{new pair beam} of the spin current contribution 
derived  for any ion velocity.
In terms of two-component spinors, the axial vector matrix elements
of electrons in ions have the form,
$(0, 2 \vec{S}_e)$ in the 4-vector notation, while
the vector matrix elements have
$(1, \epsilon_{eg} \vec{r}_{eg})$ with $\vec{r}_{eg}$ the transition dipole
moment $\vec{d}_{eg}$ divided by the electric charge, although the time
component is usually small due to the orthogonality
of wave functions.
The vector current contribution is thus obtained by
a simple replacement from the axial vector contribution:
\begin{eqnarray}
&&
4 \vec{S}_e^2 \frac{1}{\gamma} (1 + \frac{2}{3} \beta^2 \gamma^2) 
\rightarrow
\frac{4}{3} \gamma \epsilon_{eg}^2 \vec{r}_{eg}^2 = \frac{4}{3}
\gamma \epsilon_{eg}^2 \frac{\vec{d}_{eg}^2}{4\pi \alpha}
\,.
\end{eqnarray}
We have replaced the flavor component fraction at
production by a simplified result of $\sin^2 \theta_w = 1/4$,
namely $(c_b^V)^2 = (1,0,0)$.

The differential energy spectrum of a single detected neutrino
at the forward direction
is, in the high energy limit of $\gamma \gg 1$,
\begin{eqnarray}
&&
\frac{ d^2 \Gamma}{dy d\varphi} = \frac{8}{27\sqrt{\pi}(2\pi)^4} 
 N \sqrt{\rho \epsilon_{eg}} \frac{G_F^2 \epsilon_{eg}^5}{\alpha}
\gamma^{11/2} \varphi
\int dy_2  y^2 y_2 (y + y_2)^{-1/4}
\left(
4 \gamma^2 - y -y_2 - \gamma^2 y \varphi^2
\right)^{3/4}
\,,
\label {double 2n rate}
\\ &&
\frac{8}{27\sqrt{\pi}(2\pi)^4} 
\sqrt{\rho \epsilon_{eg}} \frac{G_F^2 \epsilon_{eg}^5}{\alpha}
\gamma^{11/2} \sim 
7.1 \times 10^{10}~{\rm Hz} \,
\left(\frac{ \rho \epsilon_{eg}}{ 10^{14} } \right)^{1/2} 
\left(\frac{\epsilon_{eg} }{ {\rm 10 keV} } \right)^{5} (\frac{\gamma }{10^3 })^{11/2} 
\,,  \hspace{0.5cm}
y = \frac{E}{\epsilon_{eg}} \sqrt{ \frac{1- \beta}{ 1+\beta}}
\,.~~~
\end{eqnarray}
Here $N$ is the available number of ions,
$\rho$ is the radius of the ring, $\varphi$ is the effective angle
of the neutrino pair beam.

\begin{figure*}[htbp]
 \begin{center}
   \includegraphics[width=10cm]{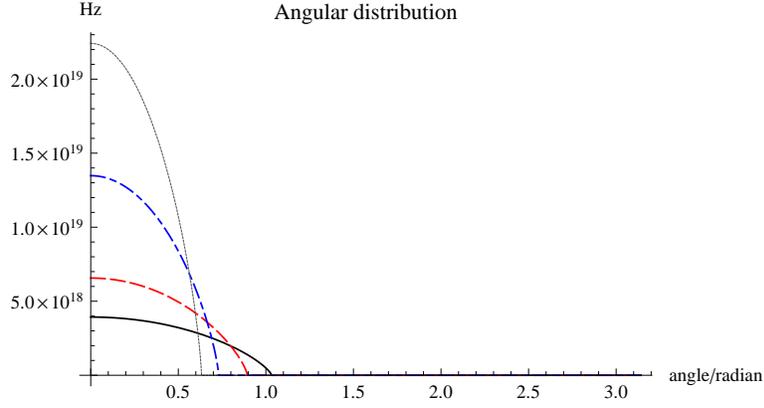}%
   \caption{
Angular distribution of single neutrino for $\gamma = 2000,
\rho \epsilon_{eg} = 10^{14}, N = 10^8, \epsilon_{eg} = 10$ keV:
neutrino energy 150 MeV in solid black,
200 MeV in dashed red, 300 MeV in dash-dotted blue,
and 400 MeV in dotted black.
}
   \label {nu angular distribution}
 \end{center} 
\end{figure*}

The angular distribution is readily calculable, and
is plotted in Fig.~\ref{nu angular distribution}.
The forward production rates
are the most relevant to neutrino oscillation
experiments away from the ring.
The forward rate is estimated by taking the angular area $\pi/\gamma^2$
times the right hand side of eq.~(\ref{double 2n rate}).
The following figure Fig.~\ref{nu energy spectrum} illustrates these rates.
The forward rates scale with ion parameters
$\propto A_{eg}\epsilon_{eg}^{5.5}$,
and with the boost factor $\propto \gamma^{3.5}$.
In order to detect $\nu_{\mu}$ events,
neutrino energies larger than 200 MeV
are desired, which gives a constraint on $2\gamma \epsilon_{eg}$.

\begin{figure*}[htbp]
 \begin{center}
  \includegraphics[width=10cm]{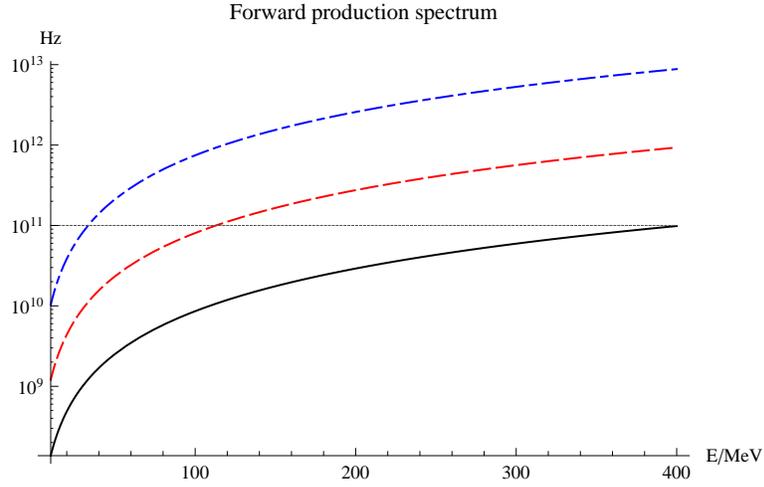}%
   \caption{
Neutrino energy spectrum rate at the forward direction of solid angle area $\pi/\gamma^2$.
Assumed parameters are $\rho \epsilon_{eg} = 10^{14}, N = 10^8$ and $\epsilon_{eg}=$10 keV,
$\gamma =$ 500 in solid black, 1000 in dashed red,
2000 in dash-dotted blue.
For reference the rate of $10^{11}$Hz is shown by the straight line.
}
   \label {nu energy spectrum}
 \end{center} 
\end{figure*}

{\bf Detection rates of neutrino events away from ion ring}
\hspace{0.3cm}
We next estimate single neutrino  event rates  by a detector 
 placed at 50 km away from the ion ring.
10 kt class of 
 $^{56}_{26}$Fe target is considered.
We shall use PDG compilation \cite{pdg} of neutrino quasi-elastic cross section extrapolated to
lower neutrino energy region.
Event rates for ideal detectors, namely the active target volume under
full coverage by the neutrino pair beam,
are
\begin{eqnarray}
&&
\nu_{\mu}({\rm Fe}); \hspace{0.3cm} 6.2 \times 10^{-11}
 P_{\mu} 
\left( \frac{E}{1~{\rm GeV}} \right)
\left( \frac{T}{{\rm 10~kt}} \right) \,
\frac{d\Gamma}{dE} \Delta E \,{\rm Hz}
\,,
\\ &&
\bar{\nu}_{\mu}({\rm Fe}); \hspace{0.3cm} 1.8 \times 10^{-11} 
P_{\bar \mu} 
\left( \frac{E}{1~{\rm GeV}} \right)
\left( \frac{T}{{\rm 10~kt}} \right)
\frac{d\Gamma}{dE} \Delta E \,{\rm Hz}
\,.
\end{eqnarray}
Thus, if the production rate $ \frac{d\Gamma}{dE} \Delta E$
presented in the previous section
is larger than $10^{11}$ Hz, then events rates at experimental sites are larger than
1 Hz.
$\Delta E$ is the energy bin taken at each energy.
Prospects for high sensitivity experiments are bright.

\vspace{0.5cm}
In summary,
we demonstrated that CPV parameter determination 
 with a high precision is possible in a short baseline
experiment using the neutrino pair beam.
Clearly, both experimental R and D works of quantum
coherent ion circulation and theoretical studies
of candidate ions are required for
further development of this new project.

\vspace{0.5cm}
 {\bf Acknowledgements}

 One of us (M.Y.) should like to thank N. Sasao for discussions on
 experimental aspects of this work.  This research was partially
 supported by Grant-in-Aid for Scientific Research on Innovative Areas
 "Extreme quantum world opened up by atoms" (21104002) from the
 Ministry of Education, Culture, Sports, Science, and Technology, and
 JSPS KAKENHI Grant Numbers 15H01031(T.A.), 15H02093 (M.T. and M.Y.), 
  25400249 (T.A.), 25400257 (M.T.), and 26105508 (T.A.).


\begin{thebibliography}{99}
\bibitem{sign of earth matter effect}
We correct the sign mistake of earth matter
effect in \cite{wolfenstein}
and \cite{barger etc}.
\end{thebibliography}
\end{document}